# A Cryogenic Integrated Noise Calibration and Coupler Module using a MMIC LNA

Eric W. Bryerton, *Senior Member, IEEE*

*Abstract*—A new cryogenic noise calibration source for radio astronomy receivers is presented. Dissipated power is only 4.2 mW, allowing it to be integrated with the cold part of the receiver. Measured long-term stability, sensitivity to bias voltages, and noise power output versus frequency are presented. The measured noise output versus frequency is compared to a warm noise diode injected into cryogenic K-band receiver and shows the integrated noise module to have less frequency structure, which will result in more accurate astronomical flux calibrations. It is currently in operation on the new 7-element K-band focal plane array receiver on the NRAO Robert C. Byrd Green Bank Telescope (GBT).

*Index Terms*—calibration, noise generators, noise measurement, radio astronomy

## I. INTRODUCTION

INTENSITY flux calibration on cryogenic radio astronomy receivers has traditionally been performed using noise diodes placed outside the cryostat and routed into the cryostat and injected into the signal path between the antenna feed and the cryogenic low-noise amplifier (LNA) through a coupler (typically ~30 dB) [1,2]. The noise diode is not integrated with the cold receiver since the power dissipation of a typical diode is typically several hundred mW or more. Fig. 1 shows a block diagram of such a typical dual-polarization radio astronomy receiver. The orthomode transducer (OMT) separates the two orthogonal polarizations received by the feed horn. Each polarization then has a noise signal injected by a coupler before proceeding through an isolator to a low-noise amplifier (LNA) and mixer. While observing, the noise diode is turned on and off with typically a 1 sec period and 50% duty cycle. The noise level is designed to be roughly 5-10% of the total system noise temperature to avoid degrading the sensitivity of the observations.

For a focal plane array receiver such as the GBT K-band focal plane array (KFPA) [3], this method of noise injection vastly complicates the cable and waveguide routing inside the cryostat as well as adds extra dewar feedthrough transitions. It is highly desirable to integrate the noise generator with the coupler on the cold stage, eliminating all the associated cabling, vacuum feedthroughs, and thermal transitions. This also has the significant benefit of less frequency structure due to standing waves and therefore more accurate calibrations. Flat spectral baselines are critical in many observations, such as detecting very faint, broad lines like CO from high-redshift galaxies [4]. It is especially important if one is trying to conclusively detect complex prebiotic molecules with more complicated spectral signatures. Using the cryogenic noise calibration module (NCM) described in this work, the typical dual-polarization radio astronomy receiver simplifies to the block diagram shown in Fig. 2. To be integrated with the cold receiver components, the noise source must be able to generate sufficient noise power while dissipating relatively little power, less than 10 mW, especially for a multi-pixel array with one noise source per pixel per polarization.

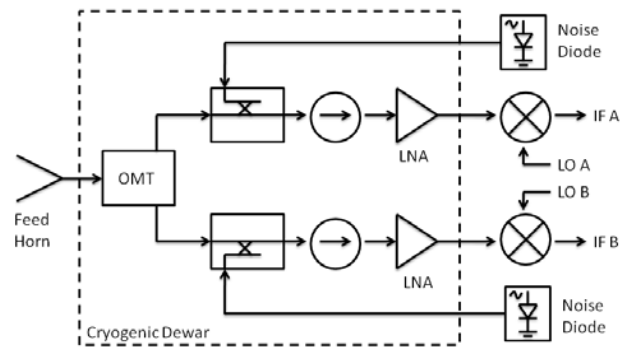

Fig. 1. Block diagram of the front end of a typical dual-polarization radio astronomy receiver.

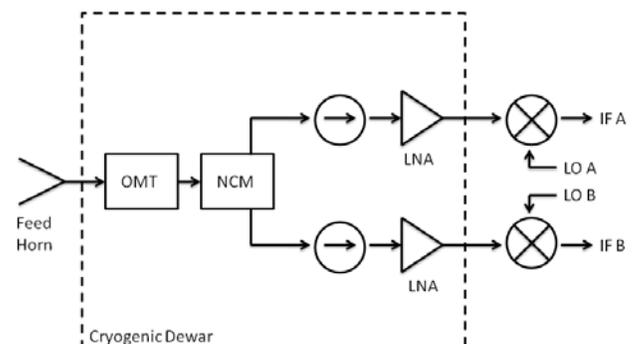

Fig. 2 Block diagram of dual-polarization radio astronomy receiver using the cryogenic noise calibration module.

This work was supported by the National Radio Astronomy Observatory, which is a facility of the National Science Foundation operated under cooperative agreement by Associated Universities, Inc.
E. W. Bryerton is with the National Radio Astronomy Observatory, Charlottesville, VA 22901 USA (phone: 434-296-0336; fax: 434-296-0324; e-mail: ebryerto@nrao.edu).

This paper describes the design, construction, and testing of a cryogenic Noise Calibration Module (NCM) built with a commercial MMIC LNA and used in a 7-pixel K-band array receiver. This noise source produces a calibration signal at the appropriate power level, dissipates only 4.2 mW, and results in smoother spectral baselines than an injected warm noise diode.

## II. Summary of Typical Astronomical Calibration

The goal of astronomical flux calibration is to translate measured power levels on a given spectrometer channel output to a source spectral flux density of the observed patch of sky ($S_{src}$). The first step is to convert a measured power level, P, into a noise temperature, T. The output power from a receiver is:

$$P = kG_{rec}T_{sys}B \qquad (1)$$

where k is Boltzmann's constant, $G_{rec}$ is receiver gain, and B is receiver bandwidth.

To remove the k, $G_{rec}$, and B factors, it is convenient to measure power ratios, i.e.

$$P_{sys,1}/P_{sys,2} = T_{sys,1}/T_{sys,2} \qquad (2)$$

To determine the noise temperature when pointed at an astronomical source ($T_{src}$) for which we are trying to determine its flux density ($S_{src}$), noise power when pointed at a nearby reference "blank" patch of sky ($P_{ref}$) is measured. The reference patch is assumed to be nearby so that the differences in noise contributed by the atmosphere are small. Then, the following power ratio is calculated:

$$P1 = (P_{src}-P_{ref})/(P_{ref}) = (T_{src}-T_{ref})/(T_{ref}) \qquad (3)$$

The desired value is $T_{src}-T_{ref}$, so (3) must be multiplied by $T_{ref}$. $T_{ref}$ cannot however be obtained by a single measurement of power, since that will contain the kGB component, so some other method to determine $T_{ref}$ is needed. This is the function of the calibration noise diode or noise calibration module (NCM). While observing the nearby reference "blank" patch of sky, the NCM is switched on and off, and the following power ratio is calculated:

$$P2 = P_{ref,off}/(P_{ref,on}-P_{ref,off}) = T_{ref,off}/(T_{ref,on}-T_{ref,off}) \qquad (4)$$

Note that $T_{ref,on} - T_{ref,off}$ is simply the noise added by the noise calibration module, $T_{cal}$, so that:

$$T_{ref} = [P_{ref,off}/(P_{ref,on}-P_{ref,off})] * T_{cal} \qquad (5)$$

Substituting (5) into (3):

$$T_{src}-T_{ref} = P1*P2*T_{cal} \qquad (6)$$

Therefore, $T_{src}-T_{ref}$ can be calculated by taking the product of two measured power ratios, P1 and P2, and $T_{cal}$.

With $T_{src}$ determined, $S_{src}$ can be calculated using:

$$T_{src} = [Area/2k]*S_{src}*\eta*exp(-\tau*A) \qquad (7)$$

where η is the antenna efficiency, τ is the atmospheric opacity, and A=1/sin(elevation). For the GBT, this equation is:

$$T_{src} = 2.84*S_{src}*\eta*exp(-\tau*A) \qquad (8)$$

where $T_{src}$ is in K and $S_{src}$ is the spectral flux density of the source in Jy ($10^{-26}$ W/m$^2$Hz). Note that to calculate $S_{src}$ from $T_{src}$, we need to know the antenna efficiency, η. For a focal plane array, η will be pixel-dependent.

$T_{cal}$ is measured in the lab before the receiver is installed on the telescope, but to account for long-term variation, $T_{cal}$ is measured on the sky by observing astronomical sources of known flux density [5] and using Eq. (7). This is typically done every few hours during an observation and can measure $T_{cal}$ to within 1% accuracy [6], better than a typical laboratory Y-factor measurement. Since $T_{cal}$ is recalibrated every few hours during an observation, it is the stability of $T_{cal}$ over hour timescales that is the critical parameter.

## III. Design Description

The fundamental noise source is a commercial off-the-shelf (COTS) MMIC LNA from United Monolithic Semiconductor (UMS), part number CHA2092b. The MMIC LNA, though not specified by the manufacturer for cryogenic operation, generates a fairly flat noise output over 18-26 GHz at 15K ambient temperature.

A few different MMIC LNAs were packaged and tested at cryogenic temperatures before settling on this particular model. The desired quantity to maximize was the "noise generation efficiency", or the amount of thermal noise generated divided by the applied dc power to the module. The reason applied dc power is important is because this module is to be used on the cold stage of a multi-pixel array. Excessive dissipated power will increase the required cooling power of the refrigerator, potentially warming the entire cryogenic portion of the receiver and degrading the receiver noise. This noise generation efficiency was first roughly estimated to be the effective input noise temperature (as specified in the MMIC datasheet, typically as a noise figure), times the specified small-signal gain, divided by the applied dc power. Commercial MMIC datasheets typically do not give cryogenic performance data, so it was necessary to actually measure several MMICs at cryogenic temperatures (about 15K in this case) to find the MMIC with the highest cryogenic noise-generating efficiency.

The MMICs were packaged in a WR-42 test block. The output of the LNA was bonded to a 5-mil thick Alumina microstrip to WR-42 E-plane probe transition. The transition has a simulated return loss greater than 20 dB from 18-26 GHz. The MMIC input was not terminated. It was found that terminating the input with a 50Ω load had little or no effect on the total noise output. For the CHA2092b, a bias of approximately 0.7 V and 6 mA produced output noise of 1000 +/- 500K over the entire 18-26.5 GHz band at 15K ambient temperature, giving a cryogenic noise-generating efficiency of approximately 240 K/mW .

Measurements of the output noise sensitivity to both gate and drain bias were also performed. As expected, the noise output is much more sensitive to gate bias than drain bias. The measured noise power sensitivity to drain voltage was approximately 10,000 K/V. To keep bias voltage fluctuation from changing noise output by no more than 1%, or about 20K, the drain voltage bias should be kept stable to within 2 mV, or about 0.3% of the 0.7 V typical bias point. Fig. 3 shows the measured output noise for three different gate voltage / drain current levels. It shows gate voltage sensitivity to be about ten times drain voltage sensitivity, or 100,000 K/V, so gate voltage bias should be kept stable to within 0.2 mV, or about 0.04% of the typical -0.54 V gate bias point. In the actual receiver, the gate voltage bias is supplied through a 10:1 resistive voltage divider. The noise source output is also expected to be highly dependent on temperature, but since this module is to be used on a high-mass cold plate inside a cryogenic Dewar, this was not a concern.

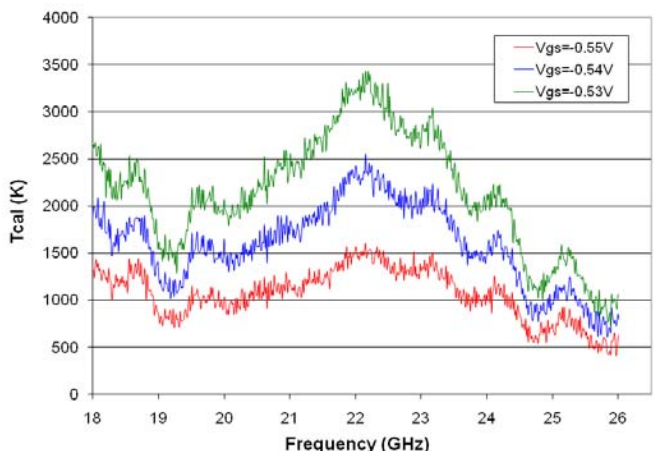

Fig. 3. Measured output noise versus frequency at 15K ambient temperature for three different bias levels. $V_{ds}$ was held constant at 0.85V.

Measurements of the long-term stability of the noise output were also performed. Fig. 4 shows the measured output noise of the MMIC test block at 15K ambient temperature over a period of 250 minutes. The small long-term fluctuation seen is believed to be due primarily to bias voltage instability, which should be minimized using the highly regulated receiver power supplies. Since the long-term stability is the critical parameter for the noise source's intended application, the short timescale fluctuations are not a concern and are in fact a result of subtracting two successive measurements of much larger power than the ~2500K noise power contributed by the noise source under test. Indeed, the short time stability of the noise source was later confirmed by stability measurements of the entire receiver on the telescope. With the noise calibration source turned on, the Allan time was consistently measured to be about 50s.

For use in a cryogenic radio astronomy receiver with two channels (one for each sky polarization), the MMIC LNA was integrated with a six-port Bethe coupler, which injects a small amount of additional noise into each polarization channel of the receiver for calibration. The noise added to each receiver channel needs to be typically 5-10% of the total system noise, so in this case, the injected noise from the calibration module was specified to be 1.5-6.0K, implying a coupling value of approximately 25dB from the MMIC LNA output into each receiver channel.

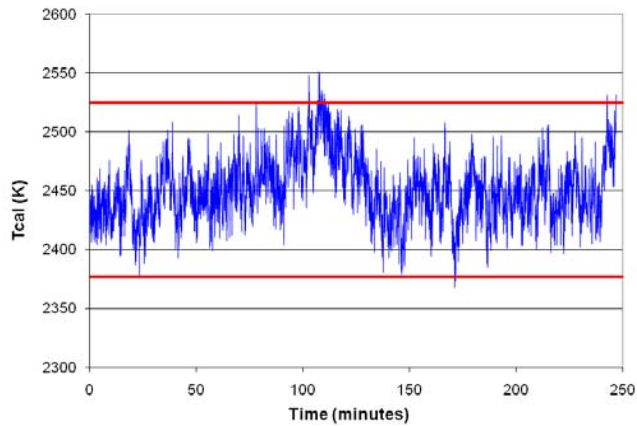

Fig. 4. Measured noise output versus time for the CHA2092b test block at 15K ambient temperature. The red lines are +/-3% away from the mean.

Photographs of the completed module are shown in Figs. 5-6. The output of the MMIC LNA enters the center waveguide channel. The two inputs from the OMT are on the left. The center channel is terminated with a conical waveguide load machined from Emerson and Cuming MF-116 ferrite absorber to absorb the uncoupled noise power from the MMIC LNA. Simulation results for the coupler show that from 18-26.5 GHz, the coupling is 25 +/- 1 dB. The input match presented to the noise source and main signal paths is better than -40 dB, the directivity is greater than 20 dB, and the RF channel isolation is better than 60dB. The coupler is fabricated by drilling holes from the outside of the block, as described in [7].

### IV. SINGLE PIXEL RECEIVER RESULTS

The return loss and insertion loss of the K-band noise calibration module (NCM) were measured with a network analyzer. Return loss measurements are shown in Fig. 7 for each channel. The return loss of a matched K-band load was also measured and is shown to indicate the quality of the network analyzer calibration. As shown, the return loss for each channel is better than 20dB across the entire band. The measured insertion loss of both channels is shown in Fig. 8. This is measured at room temperature. At cryogenic operating temperature, the insertion loss should be even less.

This NCM was then integrated into the single-pixel KFPA receiver. Receiver noise of the entire single pixel receiver was measured in the laboratory as well as on the telescope. The NCM is biased with a constant voltage supply, so that the gate voltage, $V_{gs}$, stays constant while the drain voltage, $V_{ds}$, switches from zero to its nominal value (0.667 V in this case). The IF power was recorded with a spectrum analyzer at four different LO frequencies. For each LO frequency, a Y-factor measurement was performed by measuring the IF output power while the receiver is looking into an emissive target at either room or cryogenic temperature, called the hot- and cold-loads, respectively. This Y-factor measurement is performed twice, once with the calibration noise source turned on and once with it turned off, to determine the receiver's equivalent noise temperature in both states. The difference in these noise temperatures is the injected noise, or $T_{cal}$, and is plotted in Figs. 9 and 10. The rms detector of the spectrum analyzer was used with a resolution bandwidth of 3 MHz and 0.2 s sweep time with 501 points across a 2 GHz span. Averaging was turned on with ten sweeps averaged. Therefore, the calculated accuracy of the measured power at each 4 MHz point is 1.1%. Ten successive points were averaged to give power every 40 MHz with 0.3% accuracy. The accuracy in each Y-factor calculation is then 0.6%. For the typical Y factor of 3 measured for this receiver, this gives an absolute accuracy of 0.18 for each Y factor measurement (with cal source on and cal source off). This results in a receiver noise temperature accuracy of 3%, or about +/- 0.5K for the 15-20K receiver temperature. Subtracting two receiver temperature measurements to calculate the contributed noise temperature of the calibration source gives an accuracy of +/- 1K for $T_{cal}$. This +/-1K variation is seen in Figs. 9 and 10 and is not a characteristic of the noise source. The measured $T_{cal}$ is very similar for both the LCP and RCP channels, as expected from the symmetry of the six-port Bethe coupler. Note that $V_{ds}$ remained constant for all four LO frequencies. In practice, $V_{ds}$ can be adjusted as a function of LO frequency to maintain a more constant $T_{cal}$ value. Two or three values of $V_{ds}$ need to be characterized to accomplish this. Having multiple $T_{cal}$ levels would also be useful for certain types of astronomical observations that may require specialized calibrations.

Fig. 11 shows the measured $T_{cal}$ of the single pixel test receiver compared to the measured $T_{cal}$ for the old GBT K-band receiver (K0). The new $T_{cal}$ values were averaged over 200 MHz in order to make a better comparison with the K0 values. The K0 receiver has a traditional noise calibration architecture, where a noise diode outside the Dewar at room temperature is injected into the signal path via a cold coupler inside the Dewar. As shown, the spectral structure of the new $T_{cal}$ is much smoother. The structure in the K0 calibration signal is likely not from the noise diode, but from the thermal transition, coaxial-to-waveguide transitions, and cable length between the noise diode and the cryogenic coupler. Also shown is a fourth degree polynomial fit to the new $T_{cal}$ showing how easily the $T_{cal}$ spectrum can be modeled when it is this smooth. The standard deviation of the difference between the actual measurement and the polynomial fit is 7.5% of the $T_{cal}$ value. This 7.5% is the error one would have in using this polynomial model of $T_{cal}$ rather than using an astronomical calibration to determine $T_{cal}$ versus frequency.

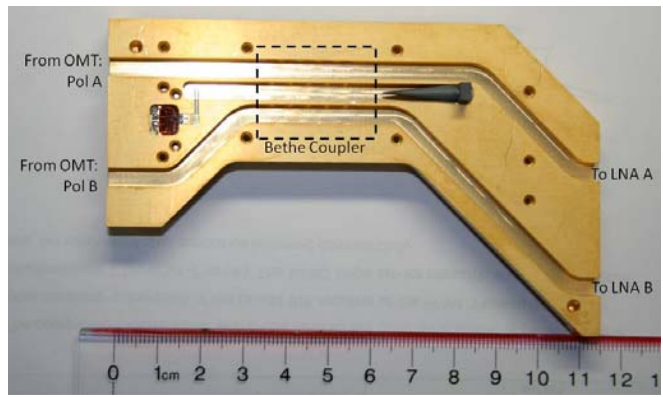

Fig. 5. Photograph of K-band noise calibration module with lid off showing placement of MMIC LNA and waveguide termination and path of waveguide channels.

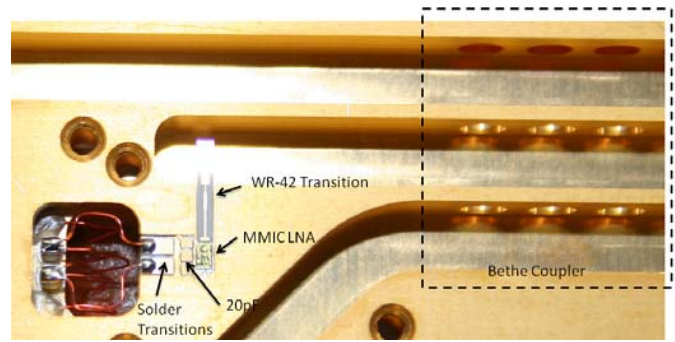

Fig. 6. Photograph of noise calibration module showing placement of MMIC LNA and Bethe coupler.

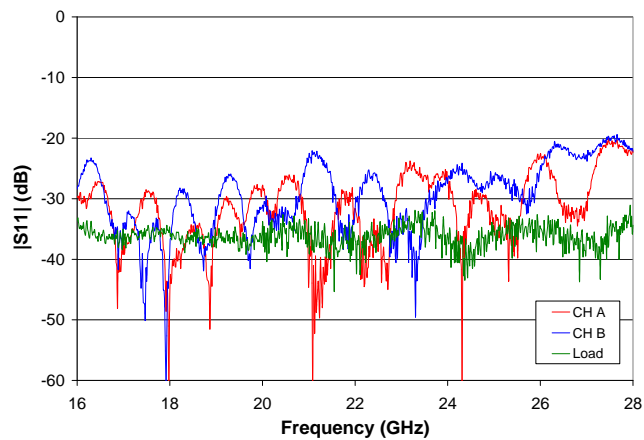

Fig. 7. Measured return loss (|s11|) versus frequency of K-band noise calibration module for both input channels as compared to waveguide load.

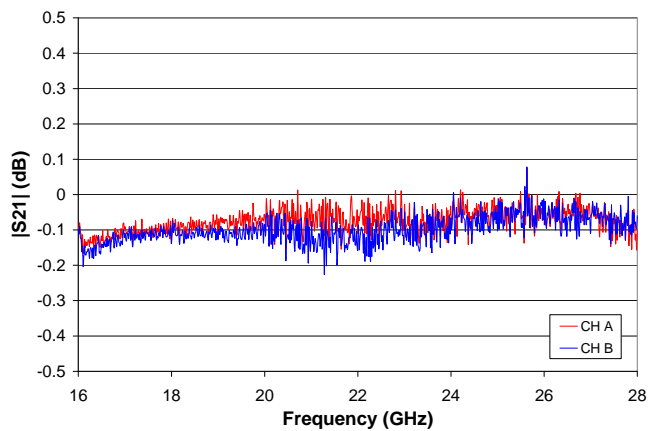

Fig. 8. Measured insertion loss (|s21|) versus frequency of K-band noise calibration module for both channels.

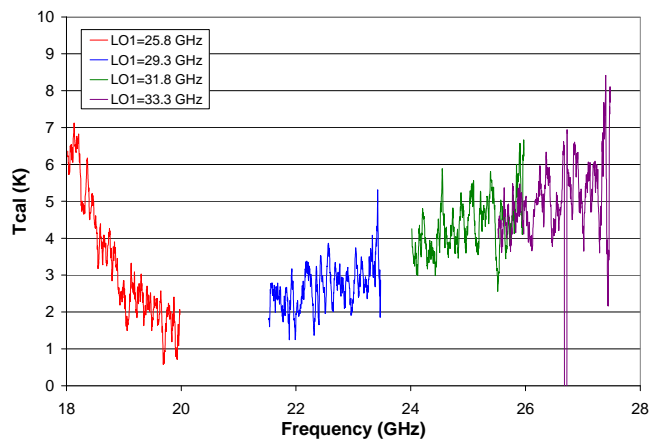

Fig. 9. Measured injected noise versus frequency for four different LO frequencies in K-band single-pixel receiver of K-band noise calibration module for the LCP (left-hand circular polarization) channel.

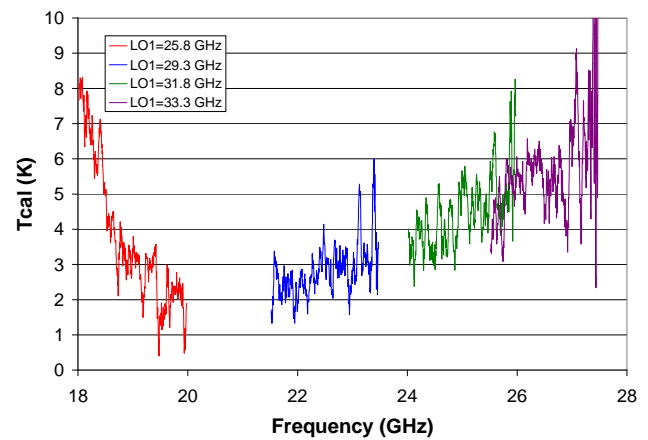

Fig. 10. Measured injected noise versus frequency for four different LO frequencies in K-band single-pixel receiver of K-band noise calibration module for the RCP (right-hand circular polarization) channel.

$T_{cal}$ was measured several times in the lab over a period of one month and was stable within a few percent. The single pixel test receiver was then placed on the telescope and successfully used to make calibrated astronomical measurements.

## V. SEVEN PIXEL ARRAY RESULTS

Eight more K-band noise calibration modules were constructed for use in the GBT K-band 7-element focal plane array. Fig. 12 shows the measured injected noise at channel B of all eight modules plus channel A of one of the modules, indicating a high level of repeatability in the noise spectrum. Since each channel of each pixel can be independently calibrated, it is not critical that these values lie on top of one another, but only that they are all relatively smooth and can be electronically tuned to give the correct range of injected noise values.

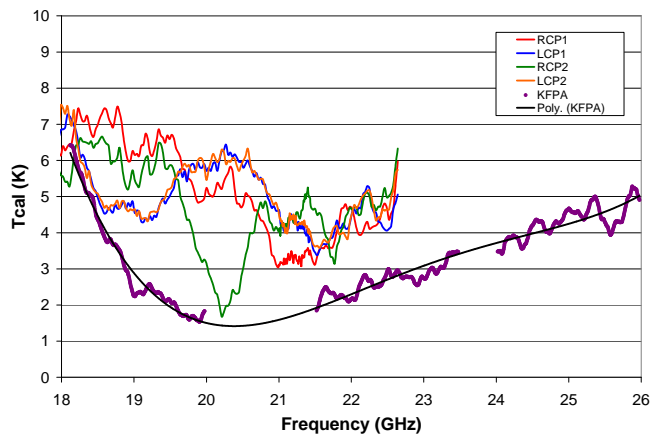

Fig. 11. Measured injected noise versus frequency of K-band noise calibration module as compared to measured injected noise of current K-band receiver using warm noise diode as noise source.

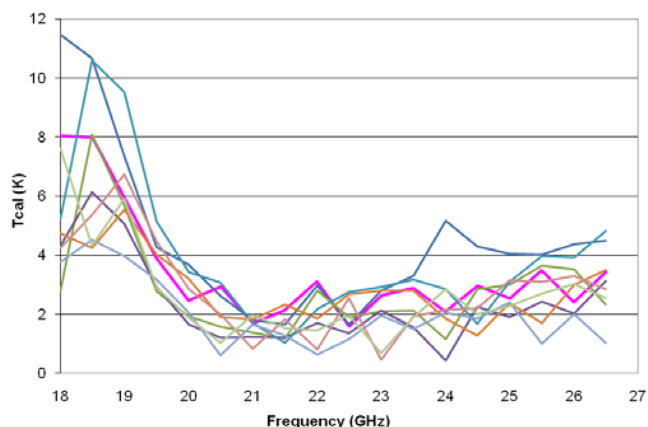

Fig. 12. Measured injected noise versus frequency at 15K ambient temperature for channel B of all 8 production modules and channel A of one module.

Fig. 13 shows the cryogenic portion of the 7-pixel K-band array receiver, indicating the placement of the NCMs. Note that the odd shape of the NCMs is to ensure closest packing of the feed horns.

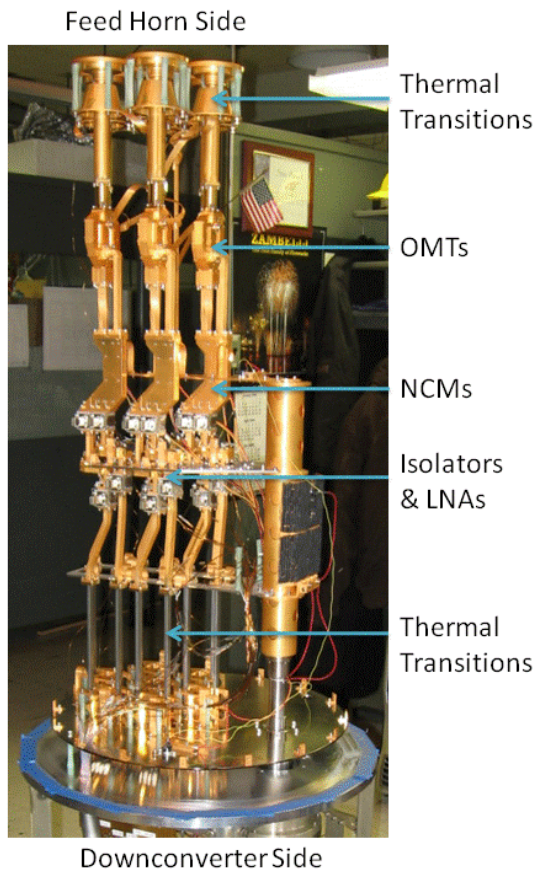

Fig. 13. Photograph of cryogenic portion of 7-element K-band array receiver showing location of noise calibration modules.

## VI. Conclusion

This paper describes the design and performance of an integrated cryogenic noise source and coupler used to provide a stable noise source with low power dissipation for astronomical receiver calibration, particularly well-suited for array receivers, where excess cabling and thermal transitions are eliminated. A prototype unit was characterized for return loss, insertion loss, noise output, and noise output stability as part of a single pixel test receiver. Several more modules were produced and integrated into a 7-pixel array receiver.

An interesting outgrowth of this work is the possibility of using a similar module for room temperature Y-factor measurements. As presented, the output noise is highly dependent on the drain current. In a well-regulated thermal environment with a well-regulated power supply, it may be possible to generate two or more well-defined and repeatable noise spectra over a certain bandwidth, which can then be used as the "hot" and "cold" loads for a Y-factor measurement. Multiple noise output levels would allow for different ratios of "hot" to "cold" noise powers, chosen to best suit the expected noise temperature of the device under test.

## Acknowledgment

The author wishes to acknowledge Tod Boyd for the microassembly of the noise calibration module; and Matt Morgan, Roger Norrod and Steve White for receiver noise measurements and helpful discussions.

**Eric W. Bryerton** (S'95–M'99–SM'06) was born in Chicago, IL on May 30, 1974. Eric received his B.S. in electrical engineering from the University of Illinois at Urbana-Champaign in 1995 and his M.S. and Ph.D. in electrical engineering from the University of Colorado at Boulder in 1997 and 1999.

He has been with the National Radio Astronomy Observatory (NRAO) Central Development Laboratory (CDL) in Charlottesville, VA since 1999. At the NRAO, he has been responsible for the design, development, and construction of the local oscillator subsystem for the Atacama Large Millimeter Array (ALMA), currently under construction in the Atacama Desert in Chile. Research interests include ultra-low-noise amplifiers and mixers for microwave through sub-millimeter-wave radio astronomy receivers and techniques for developing large format heterodyne receiver arrays.